\documentclass[12pt]{cernrep}
\usepackage{graphics}
\usepackage{graphicx}
\usepackage{epsfig}
\usepackage{amssymb}
\usepackage{amsmath}

\begin{document}

\title{CompHEP 4.5 Status Report}
\author{
E.~Boos$^1$, 
V.~Bunichev$^1$, 
M.~Dubinin$^1$, 
L.~Dudko$^1$, 
V.~Edneral$^1$, 
V.~Ilyin$^1$, 
A.~Kryukov$^1$, 
V.~Savrin$^1$, 
A.~Semenov$^2$, 
A.~Sherstnev$^3$\footnote{Contact email: a.sherstnev1@physics.ox.co.uk} \\
$^1$SINP, Moscow State University, Russia\\
$^2$JINR, Dubna, Russia \\
$^3$Rudolf Peierls Centre for Theoretical Physics, University of Oxford, 
    UK \\and SINP, Moscow State University, Russia (on leave) 
}
\maketitle
\begin{abstract}
We present a new version of the CompHEP program package, version 4.5. We 
describe new options and techniques implemented in the version: interfaces 
to ROOT and HERWIG, parallel calculations, generation of the XML-based 
header in event files (HepML), full implementation of the Les Houches 
agreements (LHA I, SUSY LHA, LHA PDF, Les Houches Event format), cascade 
matching for intermediate scalar resonances, etc. 
\end{abstract}

\section{Introduction}
\begin{figure}[hbtp]
\begin{center}
\includegraphics[width=140mm,height=100mm]{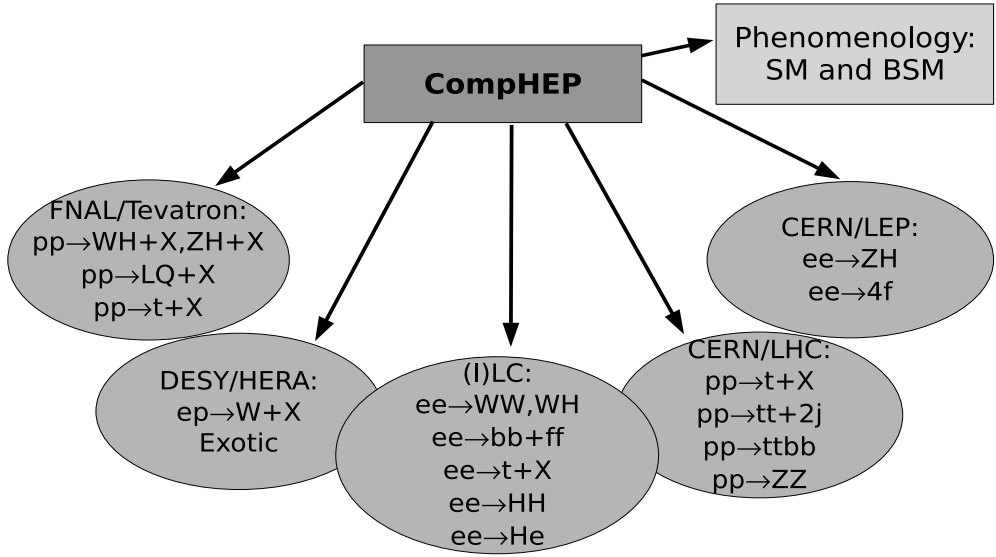}
\end{center}
\vspace{-8mm}
\caption{
Incomplete list of processes computed by means of CompHEP in recent years.}
\label{fig:past}
\end{figure}

In the last two decades ideology of automatic construction and calculation 
of tree level Feynman diagrams has proved utility and efficiency in high 
energy physics. Several codes are based on the ideology: CompHEP, 
GRACE~\cite{Fujimoto:2002sj}, MadGraph~\cite{Alwall:2007st}, 
Sherpa/Amegic~\cite{Gleisberg:2003xi}, O'Mega/Whizard~\cite{Moretti:2001zz}. 
For example, CompHEP has been applied in dozens of experimental analyses 
and hundreds of phenomenological works (see Fig.~\ref{fig:past}). The main 
practical idea behind CompHEP is to make calculations and data manipulations 
from Lagrangians, especially Beyond Standard Model Lagrangians, to final 
distributions and events with a high level of automation. Besides the 
functionality CompHEP provides a unique opportunity to get a symbolic 
answer for a matrix element squared (ME) which is specially useful in analysis 
of processes in Beyond Standard Model (BSM) scenarios. 

Current generation of experiments in high energy physics does require 
theoretical investigation of the processes with 4, 5, 6 and even 8 fermions 
and additional photons and/or gluons (jets) in the final state, for 
example, the single top production in the t-channel at hadron colliders 
($pp\to l,\nu_l,b\bar{b},j$) has 5 fermions, the top pair production with 
decays ($pp\to l,\nu_l,b\bar{b},j,j$) has 6 partons, the process 
($pp\to W^-W^++jj$) being interesting to study in the strongly interacting 
Higgs sector has 6 fermions, the process $pp\to tt+H$ for Yukawa coupling 
study leads to 8 fermions in the final state. It means we must calculate 
hundreds of extremely complicated MEs, it is impossible to do by hand. Due 
to high level of automatisation CompHEP is an indispensable tool in 
calculations of this sort. 

The previous version of CompHEP is described in~\cite{Boos:2004kh}, and a 
rather obsolete but still useful manual is~\cite{Pukhov:1999gg}. The main 
source from where users can download the code is~\cite{homepage}. 

\section{CompHEP in nutshell}
CompHEP follows the standard free software installation procedure -- {\bf 
./configure; make; make setup}. The last instruction creates a user 
working area (User Area) with scripts for CompHEP programs and some 
directories for results. Almost all operations in CompHEP a user (hereafter, 
User) can done via graphic user interface (GUI). 
By default, CompHEP uses internal parton density functions (PDFs), and it 
does not have interfaces to ROOT and XML libraries. The script {\bf 
./configure} has special options to enable ROOT~\cite{Brun:1997pa}, 
LHAPDF~\cite{Whalley:2005nh}, XML code generation, and execution 
optimisation (see {\bf ./configure --help} for more details). We test 
CompHEP on Linux (Fedora Core and Ubuntu) with gcc-3.X (by default) and 
gcc-4.X ({\bf --with-gcc4} should be used in {\bf ./configure}). Some tests 
have been done on Mac OS. 

CompHEP is divided into two separate parts, symbolic and numerical ones. 
The symbolic program is compiled and kept in the installation area, the
numerical binary is built from several libraries and C code generated by 
the symbolic program. The first thing User should do is to choose a 
model. Roughly speaking CompHEP model is a Lagrangian written by means of 
Feynman rules. Now the Lagrangian can contain 0-, 1/2-, and 1-spin particles, 
spinors can be Majorana and Dirac ones. 3- and 4- vertices with fields, 
derivatives of the fields are permitted. Technically a CompHEP model consists 
of five text files (or tables), which define the Lagrangian thoroughly. 
These tables describe a set of the fundamental particles of the model 
(names, masses/widths, spin, model charges), numerical model parameters 
(particle mass/width values, couplings, mixing parameters, other numerical 
model parameters), relations between the parameters, interaction vertices, 
and ``composite'' particles (for example, proton and artificial useful 
particle combinations). Editing of the table is possible both in CompHEP GUI 
and in a usual text editor. CompHEP has several built-in models: QED, an 
effective 4-fermion Fermi model, the SM in two different gauges: unitary 
gauge and t 'Hooft-Feynman gauge, flavour simplified SM models (they are 
called \#-models), and several SUSY models. User can create a new model 
either manually (from scratch or modifying a built-in model) or by means of 
a special tool -- LanHEP~\cite{Semenov:2008jy}. This program is a part of 
the CompHEP project. It can generate all needed model files for CompHEP (and 
also in FeynArts and LaTeX formats). LanHEP has several options for 
self-checking (charge conservation, BRST invariance, etc.). Lots of different 
BSM models have been generated with LanHEP for CompHEP. 

As soon as a model is chosen User specifies a process, for example, 
$\mathrm{u,U\to e,E,m,M,G}$ or $\mathrm{p,p\to e,E,m,M,j}$. All particles 
in the process should be defined in the used model as fundamental or 
``composite'' ones.  CompHEP constructs Feynman diagrams for the process 
and User can manipulate with the diagrams. Generally, this option is 
used for exclusion of a diagram sub-set. If the process has ``composite'' 
particles several subprocesses are built with only fundamental particles 
in the initial/final states. 
Then CompHEP constructs squared Feynman diagrams. At this stage User can 
throw out some diagrams too. The rest of the diagrams are calculated 
symbolically. CompHEP can store results of the calculation in several ways: 
for further calculations in FORM, REDUCE, MATHEMATICA, and C code in order 
to perform numerical computations and  to built a Monte Carlo generator. 
Now a ``universal'' build-in symbolic calculator is used, but we are going 
to replace it by FORM in the next CompHEP version. Capacity of the current 
symbolic calculator to compute is limited by computer resources only. Now 
CompHEP can calculate tree level Feynman diagrams for processes 
$\mathrm{1\;or\;2\to N}$ with N up to 6--7. 

Output C code generated by the CompHEP after symbolic calculations is 
compiled and linked against numerical CompHEP libraries. The main way to 
receive numerical results in CompHEP is the Monte-Carlo technique. The 
numerical CompHEP program has GUI too. It is almost independent of the main 
CompHEP installation area, except PDF data, if internal PDFs are used. 
Numerical CompHEP linked against LHAPDF depends on the LHAPDF data. 

The most complicated problem in using the numerical program is 
proper adjustment of parameters of the Monte-Carlo generator. Since the 
Monte-Carlo technique applied directly for such complicated functions as 
ME is not too efficient, we apply different techniques which improve 
calculation efficiency. CompHEP uses stratified sampling and some kind of 
universal importance sampling (VEGAS~\cite{Lepage:1977sw}). Tuning of the 
part is a bit tricky and requires some experience and understanding where 
main peaks of the ME are located. Practically the tuning means defining 
clusters in the ME (``kinematics'' menu) and describing peaks in the main 
contributing diagrams (``regularization'' menu). See more details 
in~\cite{Pukhov:1999gg}. The second part of customization is to set physical 
parameters, describe the initial state (beam particles, PDF, beam energies). 
Also User can set necessary kinematic cuts, QCD scale, and some other 
parameters. 

At the next step User defines a number of iterations and a number of calls 
per iteration for numerical calculation and starts the calculation. At first 
iterations VEGAS is automatically adjusted. Since results can be unstable 
at the stage they should be cast away. As soon as calculation process 
becomes converging, i.e. numerical errors and $\chi^2$ are small, we can 
obtain an estimation of the total cross section and build kinematic 
distributions (sub-menus ``Set distributions'' and ``View distributions''). 
User may order different built-in variables (PT, Inc. mass, rapidity, and 
some others) or define his own variable via C code in userFun.c 

After performing numerical calculations CompHEP can be used as an event 
generator. It means it can generate phase space points distributed 
according to a given ME. Event generation code uses stratified sampling 
and the standard ``hit-and-miss'' von Neuman procedure. The generation 
efficiency strongly depends on quality of the previous VEGAS tuning. At 
first, the program partitions the total phase space to cubes and searches 
for maximum values of the ME in each cube. Events are generated in each 
cube separately according to the found maximum values and contribution of 
the integral over the cube in the total cross section. Events can be stored 
in three formats, although two of them are obsolete. We recommend to use 
the LHE format (seem details in Sec.~\ref{lha}).

\section{Parallel calculations}
Since in numerical calculations CompHEP treats each subprocess separately, 
calculation of a process with lots of subprocesses (as it happens usually 
in calculations for hadron colliders) can be a laborious task. In order to 
make the task simpler and enable non-GUI calculations both symbolic and 
numerical programs in CompHEP are equipped with the batch PERL scripts 
{\bf symb\_batch.pl} and {\bf num\_batch.pl} correspondingly. These scripts 
were described in detail in~\cite{Boos:2004kh}. In addition, the scripts have 
minute help (execute {\bf ./symb\_batch.pl --help} and {\bf ./num\_batch.pl 
--help}) So, here we concentrate on new features of the scripts only. 

The main goal of the {\bf symb\_batch.pl} script is to launch the CompHEP 
symbolic program in non-GUI mode. This allows User to repeat the same 
symbolic calculation in complicated cases (it avoids mistakes and saves 
time). For example, if a particular sub-set of diagrams should be calculated. 
The main improvement in the version is a possibility to calculate diagrams 
symbolically in parallel. This option is particularly useful at multi-CPU 
machines. Several  can be started on a computer with the option -mp. For 
example, {\bf ./symb\_batch.pl -mp 4} divides the full set of diagrams to 
four sub-sets with an approximately equal number diagrams per sub-set and 
starts four symbolic programs for calculation of the four diagram sub-sets 
separately. As soon as all the sub-sets are calculated the script combines 
the prepared data and creates the standard comphep data files in {\bf tmp/}. 
There are a couple of other new auxiliary options. The option {\bf -nice} 
help to adjust scheduling priority for symbolic comphep binaries. It uses 
the standard system command {\bf nice} with (usually) permitted range for 
the priority 1-19 and the default value 10. The option {\bf -show} stat 
shows the current working status of the script, i.e. how many diagrams 
have been calculated at the moment. 

The same option ({\bf -mp}) is added to {\bf num\_batch.pl}. Here several 
independent executables can calculate subprocesses separately. In each 
numerical session CompHEP stores information in a temporary directory. As 
soon as all calculations are over, {\bf ./num\_batch.pl} combines results. 
A large number of new options of the script appear, mainly for presentation 
of results and simplification of calculations for processes with lots of 
subprocesses. The options are described in detail in the script manual 
(execute {\bf ./num\_batch.pl --help}).

\section{Implementation of Les Houches Agreements}
\label{lha}
\begin{figure}[hbtp]
\begin{center}
\includegraphics[width=150mm,height=110mm]{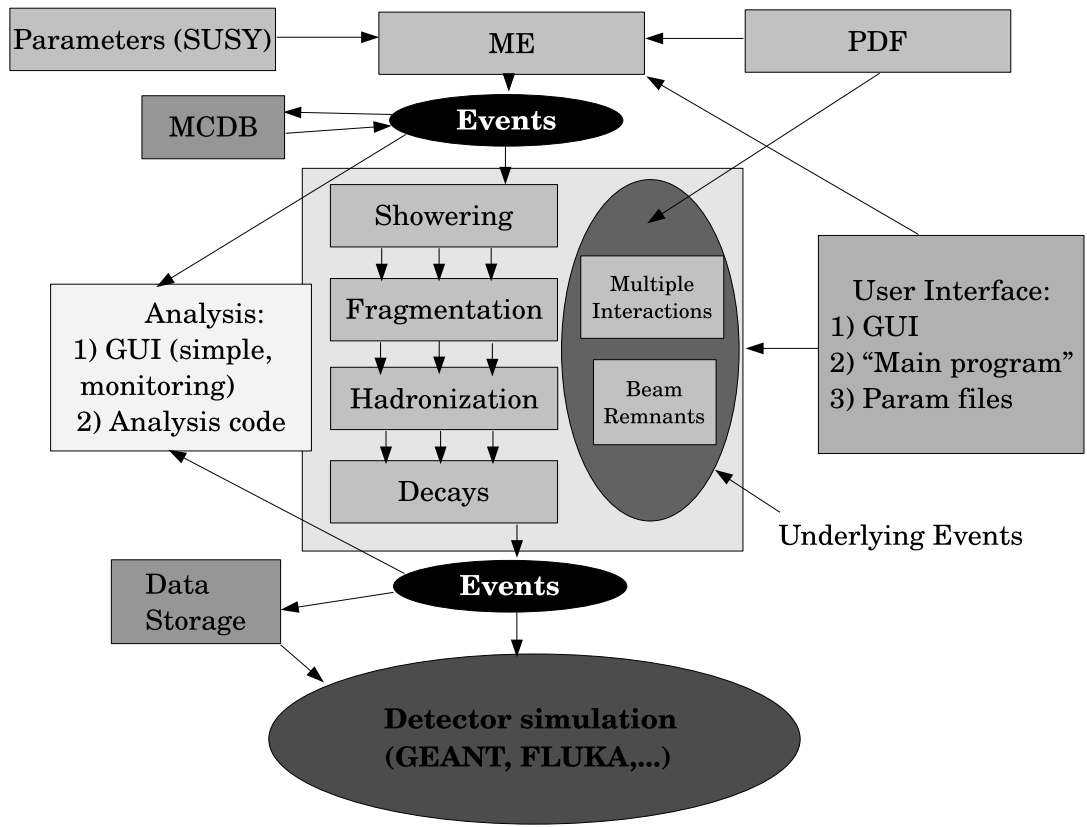}
\end{center}
\vspace{-8mm}
\caption{
A Modern picture of the Monte-Carlo event simulation chain in high energy physics
}
\label{fig:real}
\end{figure}

Fig.~\ref{fig:real} shows complexity of current Monte-Carlo generation practice. 
Due to the complexity there are many MC generators with their own advantages and 
application areas. In practical tasks we sometimes have to use several generators 
for reliable calculations. Thus, the problem of interfacing of several codes 
appears. In order to simply the problem several common ``industrial'' 
standards -- Les Houches Accords -- have been accepted. 

The first standard~\cite{Boos:2001cv}, accepted in 2001, fixes data which 
should be transferred from a Matrix Element generator (CompHEP, etc.) to 
Showering and Decay codes (PYTHIA~\cite{{Sjostrand:2006za}}, 
HERWIG~\cite{Corcella:2000bw}, etc.). Technically the accord is fixed in 
the form of FORTRAN COMMON blocks. For old CompHEP event file formats LHA 
I was implemented in CompHEP-Interfaces (see Section~\ref{interface}). 

The second Accord standardizes a presentation of parton distribution functions 
produced by different groups. PDF Les Houches Accord consists of a programming 
interface and rules applied in PDF data files. The only realization of the 
Accord has been done in the LHAPDF package~\cite{Whalley:2005nh}. Since almost 
all modern PDF sets are available in LHAPDF, interfacing CompHEP to LHAPDF 
broadens CompHEP functionality. Now CompHEP can use either internal PDFs or 
PDFs data from LHAPDF. In order to simplify installation the default option 
is internal PDFs. If User wants to use LHAPDF the package should be installed 
and CompHEP should be configured with the corresponding {\bf --with-lhapdf} 
option. After that numerical programs are linked against the LHAPDF library. 
In user interface functionality the both cases are identical. 

SUSY Les Houches Accord~\cite{Skands:2003cj} specifies a file structure for 
the parameter files of different SUSY extensions of the SM. An interface 
based on the Accord is implemented in SUGRA and GMSB SUSY models in CompHEP. 
Some external code should prepare a parameter file satisfied the Accord. 
By default, SUSPECT~\cite{Djouadi:2002ze} is used, but any other code, which 
can generate parameter files according to this Accord, can be used. CompHEP 
uses the file as input in order to apply right values of model parameters, 
mass spectrum, decay branchings. There is a script SLHA.sh in User Area 
which holds a set of initial parameters (with reasonable values) for SUSPECT. 
User can modify the parameters. The script is launched when User 
chooses a model with the interface. 

LHE is a format of event files was accepted in 2006~\cite{Alwall:2006yp}. 
In some sense it succeeds LHA I, since it performs almost the same functions 
but overcomes some practical difficulties of its predecessor, namely, 
eliminates multiplicity of files standards and adds possibility to include 
new information to event files in a standard way. CompHEP can store generated 
events in the format. Now CompHEP supports 3 event formats: cpyth1, cpyth2, 
and LHEF with  HepML header. The first two formats are obsolete ones and are 
still maintained since several experiments use the formats in their 
programming environments. We encourage users to use the LHE format. There 
is a special option ``Generator (LHEF format)'' in the event generation menu 
in {\bf n\_comphep} in order to prepare event files in the format. 

The last Les Houches Accord, BSM LHA~\cite{Alwall:2007mw}, is still being 
implemented in CompHEP.

\section{Cascade program}
Although CompHEP can generate events for processes $2\to N$ with N up to 6-7, 
complexity of calculation, both numerical and symbolic, grows drastically 
with increasing of N. But if we are interested in investigation of a new 
intermediate resonance in the process we can generate event for production 
and decay of the particle separately and combine the events. The particularly 
simplest situation happens 
if the particle is a scalar. In this case, straightforward combining gives 
right answer up to the effect of the non-zero particle width\footnote{If 
the resonance is a 1/2- or higher spin particle we have to take into account 
spin correlations.}. In many physically interesting cases zero width 
approximation is sufficient. The program cascade is created for the task. 

The program works in the following manner. There exists a script 
in User Area which starts the program. At first two event files, with 
``production'' events (where a resonance arises) and ``decay'' events 
(where the resonance decays), should be prepared.  The program takes the 
files as input and prepares a new file with combined events. 
{\bf ./cascade --help} gives details of the program syntax. 

There are two versions of the program available in CompHEP, for cpyth1 and 
for LHE. The main problem of LHE with decays is that the events can not be 
stored in the native LHE format, since it is intended to be applied for 
scattering events. It is not a problem in cpyth1. Although we have managed 
to modify original LHE for the ``decay'' events keeping the same syntax 
and re-defining sense of some parameters only, it is not accepted by other 
groups yet. So, we are going to have both versions of the code in CompHEP 
until full recognition of the modification. 

For completeness let us describe shortly the modification in LHE done for 
``decay'' events. The only problem arises in the init section, which 
describes the initial beams. In LHE it is beam particle ID and energy, and 
PDFs (for hadrons). Since we have only one particle in decays, the main 
attribute of a ``decay'' event files stored in LHE is zero ID for the second 
particle. The first particle energy is the particle mass. PDF information 
is set to zero for both beams. 

Current realization of the program has several limitations. Since events 
can have several copies of the decaying resonance cascade replaces the first 
found resonance, all other particles are left unchanged. So, if User wants 
to decay all resonances in his events he must apply the program several 
times. The most serious limitation of the current program is that cross 
sections in the output file correspond to the production events, so the 
total cross section should be multiplied by the decay branching. Since, 
generally speaking, it is not equal to 1 (if User does not include all decay 
modes in the decay event file) User should keep it mind or correct the 
cross section manually.

\section{HepML}
\label{hepml}
The current basic event format in CompHEP -- LHE -- does not describe events 
entirely. It means information inside an event file can not explain what 
kind of events the file stores, which model is used, what kinematic 
cuts are applied, etc. So, more information should be kept in event files in 
order to describe the events comprehensively. The LHE format leaves a possibility 
to keep the information. There is a place for XML tags in the file header and plain 
comment lines in event records (marked by the \# symbol). A special project has 
been started for development of XML Schema for the header. It is called HepML 
(more information is available in~\cite{hepmllinks}). Now the main XML Schema 
developed in the project is used in CompHEP and in a event storage project 
MCDB~\cite{Belov:2007qg} for parsing CompHEP event files. 

If CompHEP is linked against the libxml2 library it generates HepML code in 
LHE event files. The code can be read with a special library from the HepML 
project. The additional information includes description of a generator used, 
some information on particles in the initial and final states, description 
of cuts applied, physical parameters used (couplings, particle masses/widths, 
CKM matrix elements, etc.), beam description, some information on owners (if 
needed).

\section{ROOT functionality in CompHEP}
Histogramming is an important and useful option of CompHEP. In addition to 
the standard method to generate CompHEP histogram files ROOT histogramming 
has been added in this version. It is available via an additional sub-menu 
in the standard CompHEP histogram viewer. CompHEP creates a ROOT script with 
histogram data and a simple ROOT code for histogramming. This script, executed 
in ROOT, pictures the histogram and also creates eps and jpeg files of the 
plot. The plot contains information about the kinematic variable used and 
units of measurement. 

We introduce a new tool for superimposing ROOT-histograms in this CompHEP 
version. In the ``vegas'' integration menu we insert a new sub-menu with 
all options for combining histograms. By means of the options User can 
interactively choose some ROOT histogram files, generated by CompHEP, 
superimpose them 
on one ROOT canvas. Brief information on a selecting plot is reported, namely, 
the kinematic variable used, the variable range, and binning. The parameters 
should be the same in all combining histograms. After leaving this menu a 
ROOT file with all selected histograms is created. Curves on the plot 
have different colours. 

Events kept in any CompHEP format can be converted to a rtuple. There exists
a special auxiliary program rtupler for that. By default, CompHEP is built 
without the program, since it requires ROOT libraries. If User wants to 
have the program in User Area he should define the standard 
variable ROOTSYS (see ROOT manual) and execute {\bf ./configure} with the 
option {\bf--with-root}. After that {\bf make} builds {\bf rtupler.exe} and 
{\bf make setup} creates a shell script for launching rtupler in User Area. The 
program has a very simple syntax (see {\bf ./rtupler -h}). The output of the 
program is a rtuple and a ROOT script for a toy analysis. rtupler also copies 
two files Tchep.C and Tchep.h with the analysis code from the installation 
area. The rtuple structure is based on LHA I.

\section{CompHEP-Interfaces package}
\label{interface}
The CompHEP-Interfaces package~\cite{Belyaev:2000wn} embraces interfaces 
to showering/decay generators, namely to PYTHIA and HERWIG. Since CompHEP 
generates unweighted events on parton level for the final state, the partons 
should be translated 
to observable particles. So, PYTHIA (or HERWIG) is used for showering, 
hadronization, and decays. These interfaces are based on Les Houches 
Agreement I and needed for old CompHEP formats only. Events generated in the 
LHE format can be processed directly with internal PYTHIA/HERWIG routines. 
Although we are going to discard both old event formats (cpyth1 and cpyth2), 
these formats are used in experimental collaborations. So we shall keep the 
code in the package for the time being. Simple routines for toy analysis 
are provided both for PYTHIA and HERWIG. An interface to 
TAUOLA~\cite{Jadach:1993hs} is available by request. 
The CompHEP-Interfaces package also includes several additional programs: 
\begin{itemize}
\item[] 
{\bf mix} combines several event files to one. This utility is needed since 
CompHEP generates events in a separate file for each subprocess, but interface 
programs read events from one file only. Furthermore, by default PYTHIA read 
only one LHE event file. 

\item[] 
{\bf pdf\_reweight} substitutes parton distribution functions in an event file. 

\item[] 
{\bf make\_table/view\_table} is a part of CompHEP. The first program builds 
a kinematic distribution, with the standard CompHEP variables or any 
user-defined variable (in this case, the variable should be defined in 
userFun.c). The second program shows the distribution. 

\item[] 
{\bf translator} from cpyth1/cpyth2 to the Les Houches Event format is available 
in the package. 
As stated above CompHEP supports three different event file formats and two of 
them are obsolete. We recommend to use LHE only, since it is a common standard, 
and use old ones if User's events can not be stored in LHE properly. We plan to 
throw away the old event formats in the next CompHEP version. This program is 
available in CompHEP itself too. 

\item[] 
{\bf rtupler} is also available in the package if it is configured to use 
ROOT libraries. 
\end{itemize}

\section{Other improvements}
The new version of CompHEP allows User to build kinematic distributions in 
a user-defined rest frame of several particles. The rest frame is defined 
via a set of particle numbers in an addition column of the ``distribution'' 
table. It is called ``rest frame''. For example, the string 125 in the 
column means the rest frame with the momentum $p=p_1+p_2+p_5$. During Vegas 
integration CompHEP calculates an ordered variable and boosts it to the rest 
frame. If one leaves the field ``rest frame'' empty variables will be 
calculated in the center mass frame of the colliding beams. 

The kinematic Cut table in CompHEP has slightly been modified too. By default, 
a cut on the kinematic variable $A$ from $A_{min}$ to $A_{max}$ means keeping 
phase space points inside the interval and excluding points external to the 
interval. If User wants to exclude points inside the interval and keep outside 
points he should enter any symbol (for example, ``y'') in the last column of 
the table (``Exclusive''). It makes the cut exclusive.

\section{Final remarks}
CompHEP with the interface to PYTHIA/HERWIG is a powerful tool for 
simulation of physics at hadron and lepton colliders. CompHEP can calculate 
cross sections, build different kinematic distributions, and generate 
unweighted events at partonic level for processes $2\to N$ with N 
up to 6-7 in the Standard Model and in many extensions of the SM. CompHEP 
is compatible with modern ``Monte-Carlo industry'' standards (LHA I, SUSY 
LHA, PDF LHA, and LHE). Some improvements have been 
done in the last version, in particular, interfaces to ROOT and HERWIG, 
realisation of parallel computations both in symbolic and numerical parts 
of CompHEP. In order to facilitate interfacing of different MC codes and 
re-usage event samples CompHEP generates XML-base HepML code in event files. 

CompHEP is interfaced with software environments of experiments at the LHC 
and Tevatron. It is extensively used in various experimental analyses by 
D0, CDF, CMS, ATLAS, and ILC collaborations. 

In the next version of CompHEP 5.0 we plan to replace our built-in 
symbolic calculation with FORM computer algebra  program~\cite{Vermaseren:2000nd}. 
The code is being 
tested now and the version will be released soon. This code will allow us to 
introduce new complicated structures in the vertices, perform calculations 
with spin-2 and spin-3/2 particles, and calculations with polarised states 
by introducing the corresponding density matrices for external lines in 
diagrams.

\section{Acknowledgements}
This work was partially supported by the Russian Foundation for Basic 
Research (grants 07-07-00365-a, 08-02-91002-CERN\_a, and 08-02-92499-CNRSL\_a) 
and The Russian Ministry of Education and Science (grant NS.1685.2003.2). 
Participation of A.~Sh. is partly supported by the UK Science and Technology 
Facilities Council. 

We are grateful to our colleagues for application of CompHEP and for reporting 
problems and bugs. We thank the Organising Committee of ACAT 2008 and especially 
Federico Carminati.


\end{document}